\documentclass[11pt,thmsb,fleqn]{article}%
\usepackage{graphicx}
\usepackage{amsmath}%
\usepackage{amsfonts}%
\usepackage{amssymb}
\newtheorem{theorem}{Theorem}[section]

\newtheorem{condition}[theorem]{Condition}

\newtheorem{definition}[theorem]{Definition}

\newtheorem{notation}[theorem]{Notation}

\newtheorem{proposition}[theorem]{Proposition}

\newenvironment{proof}[1][Proof]{\textbf{#1.} }{\ \rule{0.5em}{0.5em}}
\begin{document}

\title{Multi-valued Connectives for Fuzzy Sets}
\author{Ath. Kehagias and K. Serafimidis}
\maketitle

\begin{abstract}
We present a procedure for the construction of \emph{multi-valued} t-norms and
t-conorms. Our procedure makes use of a pair of single-valued t-norms and the
respective dual t-conorms and produces interval-valued t-norms $\sqcap$ and
t-conorms $\sqcup$. In this manner we combine desirable characteristics of
different t-norms and t-conorms; if we use the t-norm $\wedge$ and t-conorm
$\vee$, then $\left(  X,\sqcap,\sqcup\right)  $ is a \emph{superlattice}, i.e.
the multivalued analog of a lattice.

\end{abstract}

\section{Introduction}

\label{sec01}

The fuzzy literature contains many examples of \emph{t-norms}, which are a
generalization of (classical) set intersection. All of these t-norms are (as
far as we know) \emph{single-valued}. To be precise: given a set (of
membership values, truth values etc.) $X$, a t-norm is a binary function
$T:X\times X\rightarrow X$ satisfying certain properties. Hence, given two
elements of $X$, call them $x,y$, then $T\left(  x,y\right)  $ is also an
element of $X$. Note that this is also true in the context of interval-valued
fuzzy sets, fuzzy sets of type 2 and other variants. For example, a t-norm
which operates on interval-valued fuzzy sets combines two intervals to produce
\emph{one} interval. Similar remarks can be made about t-conorms, which are a
generalization of (classical) set union. We will refer to both t-norms and
t-conorms as \emph{connectives}.

In this paper we introduce \emph{multi-valued} connectives. In other words, we
are interested in binary functions which map elements of $X$ to \emph{subsets}
of $X$. Before formally presenting our results let us briefly discuss the
reasons for introducing multi-valued connectives.

Fuzzy theorists have often argued that a major motive behind the theory of
fuzzy sets has been the treatment of \emph{uncertainty}. Many examples apppear
in the literature; for instance Nguyen \cite{Nguyen} mentions classes with
vaguely defined boundaries and numbers which are only known to lie within an
interval as two examples where fuzzy sets can be fruitfully applied.

The above examples (and many similar ones appearing throughout the literature)
involve uncertainty about the degree to which objects belong to sets; on the
other hand the manner in which fuzzy sets are \emph{combined }(e.g. by unions,
intersections etc.) \emph{does not involve any uncertainty}. For example,
given two fuzzy sets $A$ and $B$ and an element $x$, the degree to which $x$
belongs to both $A$ and $B$ is given by $A\left(  x\right)  \wedge B\left(
x\right)  $; no uncertainty is involved in the application of the $\wedge$
connective. A natural extension of the principle of fuzziness is to consider
\emph{uncertain connectives}; the use of multi-valued t-norms and t-conorms is
a simple step in this direction.

Hence the plan of this paper is as follows. We work in the context of a
deMorgan lattice $\left(  X,\wedge,\vee,^{\prime}\right)  $ (where $^{\prime}$
is \emph{negation}), hence our results will hold equally for fuzzy and
\emph{L-fuzzy }sets. We introduce multi-valued operations $\sqcap:X\times
X\rightarrow\mathbf{P}\left(  X\right)  $ and $\sqcup:X\times X\rightarrow
\mathbf{P}\left(  X\right)  $ (where $\mathbf{P}\left(  X\right)  $ is the
\emph{power set} of $X$). Then we show that $\sqcap$ has properties which are
analogous (in the multi-valued context) of the properties usually required of
t-norms; similarly $\sqcup$ has propeties analogous to those usually required
of t-conorms. Finally, we show that the structure $\left(  X,\sqcap
,\sqcup\right)  $ is the analog (in the multi-valued context) of a lattice.

The last remark requires some additional explanation. Let us first remark that
there is an extensive literature in the study of multi-valued algebraic
operations (called \emph{hyperoperations}) and the corresponding algebraic
structures (\emph{hyperalgebras}). The books \cite{Corsini1,Corsini2} present
an extensive study of hyperalgebras such as \emph{hypergroups} (the
multi-valued analog of group, see also \cite{Calugareanu1,Vougiouk2} and for
\emph{fuzzy }hypergroups
\cite{CorsiniTofan,Kehagias1,Kehagias3,zah11,zah12,zah01} and other relations
between hypergroups and fuzzy sets
\cite{Corsini3,CorsiniLeoreanu1,CorsiniLeoreanu2,Leoreanu1,Kehagias5}),
hyperrings (the multi-valued analog of ring, see also \cite{Spartal1,zah02}),
hyperlattices and \emph{superlattices} (the multi-valued analog of lattice,
see also
\cite{Ashrafi1,Gutan1,HyperLat1,MitKon1,MitKon11,MitKon2,MitKon12,Rah01,Rah02,Superlattice}%
) etc. As will be seen in Section \ref{sec03}, our $\left(  X,\sqcap
,\sqcup\right)  $ is a \emph{superlattice} \cite{Superlattice}.

While multi-valued operations have been studied extensively in the
hyperalgebraic literature, we believe (as already mentioned) that they have
not been previously discussed in the fuzzy literature. However, our approach
is quite similar to the one used by Jenei in \cite{Jenei1}. Jenei introduces
t-norms and t-conorms for intervals, i.e. his connectives are
\emph{single-valued} functions which map pairs of intervals to intervals.
Hence these are essentially connectives for interval-valued fuzzy sets; the
same idea is discussed in \cite{Nguyen} and several other places. However, the
actual construction of the interval-valued t-norms and t-conorms is the same
as the one used by us (indeed Jenei's paper has been a major inspiration to
us). Jenei argues that his connectives are preferrable to classical ones
because they combine a large number of desirable properties; this remark also
holds for our $\sqcap$ and $\sqcup$ and can be considered as an additional
reason for their introduction.

\section{Preliminaries}

\label{sec02}

We will present our results in the context of L-fuzzy sets, i.e. all the
results presented below hold when membership takes values in a lattice (rather
than in the unit interval of real numbers). This generality can be obtained at
no additional cost, i.e. the proofs of our results are essentially the same
for the cases of real numbers and general lattice.

Hence, in what follows we assume the existence of a deMorgan lattice $\left(
X,\wedge,\vee,^{\prime}\right)  $ (where $^{\prime}$ denotes \emph{negation})
with a minimum element 0 and a maximum element 1. The order compatible with
$\wedge,\vee$ will be denoted by $\leq$. Lattice \emph{intervals} are defined
in the standard manner: for every $x,y\in X$ with $x\leq y$ we define $\left[
x,y\right]  =\left\{  z:x\leq z\leq y\right\}  $. The \emph{empty }interval is
the empty set $\emptyset$ and can be symbolized as $\left[  x,y\right]  $ for
any pair $x,y$ such that $x\nleq y$. The collection of all intervals of $X$,
including the empty interval, will be symbolized by $\mathbf{I}\left(
X\right)  $. We define, in standard manner, an order on $\mathbf{I}\left(
X\right)  $.

\begin{definition}
\label{cnt0201}For every $\left[  x,y\right]  $, $\left[  u,v\right]
\in\mathbf{I}\left(  X\right)  $ we write $\left[  x,y\right]  \preceq\lbrack
u,v]$ iff $x\leq u$ and $y\leq v$.
\end{definition}

\begin{proposition}
\label{cnt0202}$\preceq$ is an order on $\mathbf{I}\left(  X\right)  $ and
$\left(  \mathbf{I}\left(  X\right)  ,\preceq\right)  $ is a lattice where
\[
\inf\left(  \left[  x,y\right]  ,\left[  u,v\right]  \right)  =\left[  x\wedge
y,u\wedge v\right]  ,\qquad\sup\left(  \left[  x,y\right]  ,\left[
u,v\right]  \right)  =\left[  x\vee y,u\vee v\right]
\]
for every $\left[  x,y\right]  $, $\left[  u,v\right]  \in\mathbf{I}\left(
X\right)  $.
\end{proposition}

In the lattice context we can define a t-norm $T$ $\ $to be any function
$T:X\times X\rightarrow X$ which satisfies the following properties.

\begin{definition}
\label{cnt0203}A function $T:X\times X\rightarrow X$ is a \emph{t-norm} if it
satisfies the following for every $x,y,z\in X$.

\begin{enumerate}
\item $T\left(  1,x\right)  =x.$

\item $T\left(  x,y\right)  =T\left(  y,x\right)  .$

\item $T\left(  x,T\left(  y,z\right)  \right)  =T\left(  T\left(  x,y\right)
,z\right)  .$

\item $x\leq y\Rightarrow T\left(  x,z\right)  \leq T\left(  y,z\right)  $.
\end{enumerate}
\end{definition}

Similarly, a t-conorm $S$ $\ $is any function $S:X\times X\rightarrow X$ which
satisfies the following properties.

\begin{definition}
\label{cnt0204}A function $S:X\times X\rightarrow X$ is a \emph{t-conorm} if
it satisfies the following for every $x,y,z\in X$:

\begin{enumerate}
\item $S\left(  0,x\right)  =x.$

\item $S\left(  x,y\right)  =S\left(  y,x\right)  .$

\item $S\left(  x,S\left(  y,z\right)  \right)  =S\left(  S\left(  x,y\right)
,z\right)  .$

\item $x\leq y\Rightarrow S\left(  x,z\right)  \leq S\left(  y,z\right)  $.
\end{enumerate}
\end{definition}

\begin{notation}
\label{cnt0205}We will write $T\left(  x,y,z\right)  $ for $T\left(  T\left(
x,y\right)  ,z\right)  =T\left(  x,T\left(  y,z\right)  \right)  $ and
$S\left(  x,y,z\right)  $ for $S\left(  S\left(  x,y\right)  ,z\right)
=S\left(  x,S\left(  y,z\right)  \right)  $ (by associativity).
\end{notation}

\begin{definition}
\label{cnt0206}Given a t-norm $T$ and a t-conorm $S$, we say that $T$ and $S$
are \emph{dual }(with respect to the negation $^{\prime}$) iff $\left(
T\left(  x,y\right)  \right)  ^{\prime}=S\left(  x^{\prime},y^{\prime}\right)
$.
\end{definition}

\begin{definition}
\label{cnt0207}For every $\left[  x,y\right]  \in\mathbf{I}\left(  X\right)
$, we define $\left[  x,y\right]  ^{\prime}=\left\{  z^{\prime}\right\}
_{z\in\left[  x,y\right]  }$.
\end{definition}

\textbf{Remark}. In the sequel we will occasionally make use of certain
well-known properties of t-norms and t-conorms which follow from Definitions
\ref{cnt0203} and \ref{cnt0204}. For example, $T\left(  0,x\right)  =0$,
$S\left(  1,x\right)  =1$, $x\leq y\Rightarrow$ $T\left(  z,x\right)  \leq
T\left(  z,y\right)  $, $x\leq y\Rightarrow$ $S\left(  z,x\right)  \leq
S\left(  z,y\right)  $ etc. Also, using Definition \ref{cnt0207} it is
straightforward that $\left[  x,y\right]  ^{\prime}=\left[  y^{\prime
},x^{\prime}\right]  $. Finally, proofs of the following propositions can be
found in \cite{Nguyen}.

\begin{proposition}
\label{cnt0208}$\wedge$ is a t-norm and $\vee$ is its dual t-conorm.
\end{proposition}

\begin{proposition}
\label{cnt0209}Given a t-norm $T$ and a t-conorm $S$, for every $x,y\in X$ we
have: $T\left(  x,y\right)  \leq x\wedge y$ and $x\vee y\leq S\left(
x,y\right)  $.
\end{proposition}

\begin{proposition}
\label{cnt0210}For all $x,y\in X$ we have: $T\left(  x,y\right)  \leq x$,
$x\leq S\left(  x,y\right)  $.
\end{proposition}

We now present some material relating to \emph{hyperoperations}. For more
details see \cite{Corsini1}.

\begin{definition}
\label{cnt0211}A \emph{hyperoperation }is a mapping $\ast:X\times
X\rightarrow\mathbf{P}\left(  X\right)  $, where $\mathbf{P}\left(  X\right)
$ is the power-set of $X$.
\end{definition}

\textbf{Remark}. In other words, while an operation maps every pair of
elements to an element, a hyperoperation maps every pair of elements to a
\emph{set}. The following is a standard notation used in the hyperoperations literature.

\begin{notation}
\label{cnt0212}If $\ast$ is a hyperoperation on $X$, then for every $x,y,z\in
X$ we define
\[
x\ast\left(  y\ast z\right)  =\cup_{u\in y\ast z}x\ast u,\qquad\left(  x\ast
y\right)  \ast z=\cup_{u\in x\ast y}u\ast z.
\]
\end{notation}

A particular hyperstructure of interest in this paper is the
\emph{superlattice} \cite{Jakubik1,Superlattice}.\emph{ }

\begin{definition}
\label{cnt0213}Given hyperoperations $\bigtriangledown,\bigtriangleup$ on
$\left(  X,\wedge,\vee\right)  $, we say that $\left(  X,\bigtriangledown
,\bigtriangleup\right)  $ is a \ \emph{superlattice} iff the following
properties hold for all $x,y,z\in X$.

\begin{description}
\item[A1] $x\in x\bigtriangleup x$, $x\in x\bigtriangledown x$.

\item[A2] $x\bigtriangleup y=y\bigtriangleup x$, $x\bigtriangledown
y=y\bigtriangledown x$.

\item[A3] $\left(  x\bigtriangleup y\right)  \bigtriangleup z=x\bigtriangleup
\left(  y\bigtriangleup z\right)  $, $\left(  x\bigtriangledown y\right)
\bigtriangledown z=x\bigtriangledown\left(  y\bigtriangledown z\right)  $.

\item[A4] $x\in\left(  x\bigtriangleup y\right)  \bigtriangledown x$,
$x\in\left(  x\bigtriangledown y\right)  \bigtriangleup x$.

\item[A5] $x\leq y\Leftrightarrow y\in x\bigtriangledown y$ $\Leftrightarrow
x\in x\bigtriangleup y$.
\end{description}
\end{definition}

Obviously this is a generalization of the concept of lattice to the context of
hyperoperations; in particular, every lattice can be seen as a superlattice
with ``trivial'' (single-valued) hyperoperations.

\section{Interval-Valued t-Norms and t-Conorms}

\label{sec03}

In the following $T\left(  x,y\right)  $ will denote an arbitrary t-norm and
$S\left(  x,y\right)  $ its dual t-conorm (with respect to some arbitrary
negation $x^{\prime}$). The only condition we impose on $T\left(  x,y\right)
$ and $S\left(  x,y\right)  $ is the following.

\begin{condition}
\label{cnt0302}For all $x,y,z\in X$ we have:

\begin{enumerate}
\item $T\left(  x\vee y,z\right)  =T\left(  x,z\right)  \vee T\left(
y,z\right)  .$

\item $T\left(  x\wedge y,z\right)  =T\left(  x,z\right)  \wedge T\left(
y,z\right)  .$

\item $S\left(  x\vee y,z\right)  =S\left(  x,z\right)  \vee S\left(
y,z\right)  .$

\item $S\left(  x\wedge y,z\right)  =S\left(  x,z\right)  \wedge S\left(
y,z\right)  .$
\end{enumerate}
\end{condition}

\begin{proposition}
\label{cnt0303}Condition \ref{cnt0302} is automatically satisfied for every
$T,S$ pair if $X$ is the interval $\left[  0,1\right]  $ of real numbers.
\end{proposition}

\begin{proof}
We only prove the first part of Condition \ref{cnt0302} (the remaining parts
are proved similarly). Without loss of generality suppose that $x\leq y$. Then
$x\vee y=y$ and so $T\left(  x\vee y,z\right)  =T\left(  y,z\right)  $. But
also $x\leq y\Rightarrow$ $T\left(  x,z\right)  \leq T\left(  y,z\right)
\Rightarrow$ $T\left(  x,z\right)  \vee T\left(  y,z\right)  $= $T\left(
y,z\right)  $.
\end{proof}

We now define the interval-valued fuzzy connectives $\sqcap,\sqcup$.

\begin{definition}
\label{cnt0304}For all $x,y\in X$ we define $x\sqcap y=\left[  T\left(
x,y\right)  ,x\wedge y\right]  $, $x\sqcup y=\left[  x\vee y,S\left(
x,y\right)  \right]  $.
\end{definition}

\begin{proposition}
\label{cnt0305}For all $x,y,z\in X\ $such that $y\leq z$, we have:
$x\sqcap\left[  y,z\right]  $= $\left[  T\left(  x,y\right)  ,x\wedge
z\right]  $ and $x\sqcup\left[  y,z\right]  $= $\left[  x\vee y,S\left(
x,z\right)  \right]  .$
\end{proposition}

\begin{proof}
Choose any $w\in x\sqcap\left[  y,z\right]  $= $\cup_{u\in\left[  y,z\right]
}x\sqcap u$= $\cup_{y\leq u\leq z}\left[  T\left(  x,u\right)  ,x\wedge
u\right]  $. Then there exists some $u$ such that: $y\leq u\leq z$ and
$T\left(  x,u\right)  \leq w\leq x\wedge u$. It follows that $w\leq x\wedge
u\leq x\wedge z$ and and $T\left(  x,y\right)  \leq T\left(  x,u\right)  \leq
w$. Hence $w\in\left[  T\left(  x,y\right)  ,x\wedge z\right]  $ and so
\begin{equation}
x\sqcap\left[  y,z\right]  \subseteq\left[  T\left(  x,y\right)  ,x\wedge
z\right]  . \label{eq0311}%
\end{equation}

On the other hand, choose any $w\in\left[  T\left(  x,y\right)  ,x\wedge
z\right]  $ and define
\begin{equation}
u=\left(  y\vee w\right)  \wedge z=y\vee\left(  w\wedge z\right)
\label{eq0312}%
\end{equation}
(the second equality in (\ref{eq0312}) follows from distributivity). Now
\begin{align}
u  &  =\left(  y\vee w\right)  \wedge z\leq z\label{eq0313}\\
u  &  =y\vee\left(  w\wedge z\right)  \geq y. \label{eq0314}%
\end{align}
Hence
\begin{equation}
u\in\left[  y,z\right]  \label{eq0315}%
\end{equation}
On the other hand, $u\wedge x=\left(  y\vee w\right)  \wedge z\wedge x$. But
$w\leq y\vee w\ $ and $w\leq z\wedge x$, hence $w\leq u\wedge x$. Also
$T\left(  u,x\right)  $= $T\left(  y\vee\left(  w\wedge z\right)  ,x\right)
$= $T\left(  y,x\right)  \vee T\left(  w\wedge z,x\right)  $. But $T\left(
y,x\right)  \leq w$ and $T\left(  w\wedge z,x\right)  \leq T\left(
w,x\right)  \leq w$. Hence $T\left(  u,x\right)  $= $T\left(  y\vee\left(
w\wedge z\right)  ,x\right)  \leq w$. Hence
\begin{equation}
w\in\left[  T\left(  u,x\right)  ,u\wedge x\right]  . \label{eq0316}%
\end{equation}
(\ref{eq0315}) and (\ref{eq0316}) imply that $w\in x\sqcap\left[  y,z\right]
$ and so
\begin{equation}
\left[  T\left(  x,y\right)  ,x\wedge z\right]  \subseteq x\sqcap\left[
y,z\right]  ; \label{eq0317}%
\end{equation}
(\ref{eq0311}) and (\ref{eq0317}) imply that $\left[  T\left(  x,y\right)
,x\wedge z\right]  =x\sqcap\left[  y,z\right]  $ and we have proved the first
part of the theorem; the second part is proved dually.
\end{proof}

The following proposition shows that $\sqcap,\sqcup$ have the analogs of
t-norm, t-conorm properties (in the context of hyperoperations).

\begin{proposition}
\label{cnt0306}For all $x,y,z\in X\ $we have:

\begin{enumerate}
\item $x\in1\sqcap x$, $0\in0\sqcap x$, $x\in0\sqcup x$, $1\in1\sqcup x$.

\item $x\sqcap y=y\sqcap x$, $x\sqcup y=y\sqcup x$.

\item If $x\leq y$, then $x\sqcap z\preceq y\sqcap z$ and $x\sqcup z\preceq y$
$\sqcup z$.

\item $\left(  x\sqcap y\right)  \sqcap z=x\sqcap\left(  y\sqcap z\right)
=\left[  T\left(  x,y,z\right)  ,x\wedge y\wedge z\right]  \ $and $\left(
x\sqcup y\right)  \sqcup z=x\sqcup\left(  y\sqcup z\right)  =\left[  x\vee
y\vee z,S\left(  x,y,z\right)  \right]  .$
\end{enumerate}
\end{proposition}

\begin{proof}
The first part of 1 is proved as follows: $1\sqcap x$= $\left[  T\left(
1,x\right)  ,1\wedge x\right]  $= $\left[  x,x\right]  \ni x$. Similarly, for
the second part: $0\sqcap x$= $\left[  T\left(  0,x\right)  ,0\wedge x\right]
$= $\left[  0,0\right]  \ni0$. The remaining two parts are proved similarly. 2
is immediate. Regarding 3 we have: $x\sqcap z=\left[  T\left(  x,z\right)
,x\wedge z\right]  $, $y\sqcap z=\left[  T\left(  y,z\right)  ,y\wedge
z\right]  $; now, if $x\leq y$ then $T\left(  x,z\right)  \leq T\left(
y,z\right)  $ and $x\wedge z\leq$ $y\wedge z$ which shows that $x\sqcap
z\preceq y\sqcap z$; $x\sqcup z\preceq y$ $\sqcup z$ is proved dually. Let us
now turn to 4.

First, take any $w\in\left(  x\sqcap y\right)  \sqcap z=\cup_{u\in x\sqcap
y}u\sqcap z=\cup_{T\left(  x,y\right)  \leq u\leq x\wedge y}\left[  T\left(
u,z\right)  ,u\wedge z\right]  $. Hence there exists some $u$ such that
$T\left(  x,y\right)  \leq u\leq x\wedge y$ and $T\left(  u,z\right)  \leq
w\leq u\wedge z.$ Hence $w\leq u\wedge z\leq x\wedge y\wedge z$ and $w\geq$
$T\left(  u,z\right)  \geq$ $T\left(  T\left(  x,y\right)  ,z\right)  $=
$T\left(  x,y,z\right)  $. It follows that $w\in\left[  T\left(  x,y,z\right)
,x\wedge y\wedge z\right]  $ and so
\begin{equation}
\left(  x\sqcap y\right)  \sqcap z\subseteq\left[  T\left(  x,y,z\right)
,x\wedge y\wedge z\right]  . \label{eq0301}%
\end{equation}

Second, take any $w\in\left[  T\left(  x,y,z\right)  ,x\wedge y\wedge
z\right]  $ and define
\begin{equation}
u=\left(  T\left(  x,y\right)  \vee w\right)  \wedge\left(  x\wedge y\right)
=T\left(  x,y\right)  \vee\left(  w\wedge x\wedge y\right)  \label{eq0302}%
\end{equation}
(the second equality in (\ref{eq0302}) follows from distributivity). Now
\begin{align}
u  &  =\left(  T\left(  x,y\right)  \vee w\right)  \wedge\left(  x\wedge
y\right)  \Rightarrow u\leq x\wedge y\label{eq0303}\\
u  &  =T\left(  x,y\right)  \vee\left(  w\wedge x\wedge y\right)  \Rightarrow
u\geq T\left(  x,y\right)  \label{eq0304}%
\end{align}
and hence
\begin{equation}
u\in\left[  T\left(  x,y\right)  ,x\wedge y\right]  . \label{eq0305}%
\end{equation}
Furthermore $u\wedge z=\left(  T\left(  x,y\right)  \vee w\right)
\wedge\left(  x\wedge y\right)  \wedge z$. But $w\leq T\left(  x,y\right)
\vee w$ and $w\leq\left(  x\wedge y\right)  \wedge z$. Hence $w\leq u\wedge
z.$ Also, $T\left(  u,z\right)  =T\left(  T\left(  x,y\right)  \vee\left(
w\wedge x\wedge y\right)  ,z\right)  $= $T\left(  T\left(  x,y\right)
,z\right)  \vee T\left(  w\wedge x\wedge y,z\right)  $. Now $T\left(  T\left(
x,y\right)  ,z\right)  $= $T\left(  x,y,z\right)  \leq w$ and $T\left(
w\wedge x\wedge y,z\right)  \leq$ $T\left(  w,z\right)  \leq$ $w$. Hence
$T\left(  u,z\right)  \leq w$. In short
\begin{equation}
w\in\left[  T\left(  u,z\right)  ,u\wedge z\right]  . \label{eq0306}%
\end{equation}
From (\ref{eq0305})\ and (\ref{eq0306}) we conclude $w\in\left(  x\sqcap
y\right)  \sqcap z$ and so%
\begin{equation}
\left[  T\left(  x,y,z\right)  ,x\wedge y\wedge z\right]  \subseteq\left(
x\sqcap y\right)  \sqcap z. \label{eq0307}%
\end{equation}
From (\ref{eq0301})\ and (\ref{eq0307}) we conclude $\left[  T\left(
x,y,z\right)  ,x\wedge y\wedge z\right]  =\left(  x\sqcap y\right)  \sqcap z$
and we have established the first part of 4; the second part is proved dually.
\end{proof}

\begin{proposition}
\label{cnt0307}For all $x,y\in X\ $we have:

\begin{enumerate}
\item $x\in x\sqcap x$, $x\in x\sqcup x$.

\item $x\in x\sqcap\left(  x\sqcup y\right)  $, $x\in x\sqcup\left(  x\sqcap
y\right)  $.

\item $x\leq y\Leftrightarrow y\in x\sqcup y$ $\Leftrightarrow x\in x\sqcap y$.
\end{enumerate}
\end{proposition}

\begin{proof}
For 1: $x\sqcap x=\left[  T\left(  x,x\right)  ,x\wedge x\right]  $, but
$T\left(  x,x\right)  \leq x$ and $x\wedge x=x$, hence $x\in x\sqcap x$; the
second part is proved dually.

For 2: $x\sqcap\left(  x\sqcup y\right)  $= $x\sqcap\left[  x\vee y,S\left(
x,y\right)  \right]  $= $\left[  T\left(  x,x\vee y\right)  ,x\wedge S\left(
x,y\right)  \right]  $. But $T\left(  x,x\vee y\right)  $= $T\left(
x,x\right)  \vee T\left(  x,y\right)  $ and we have $T\left(  x,x\right)  \leq
x$, $T\left(  x,y\right)  \leq x$; hence $T\left(  x,x\vee y\right)  \leq x$.
Also, $x\leq S\left(  x,y\right)  $ and so $x\wedge S\left(  x,y\right)  =x$.
Hence $x\in\left[  T\left(  x,x\vee y\right)  ,x\wedge S\left(  x,y\right)
\right]  $ = $x\sqcap\left(  x\sqcup y\right)  $. The second part of 2 is
proved dually.

Finally, for 3, $x\sqcup y=\left[  x\vee y,S\left(  x,y\right)  \right]  $,
but $x\leq y\Rightarrow$ $x\vee y=y$ and $y\leq S\left(  x,y\right)  $; hence
$y\in\left[  y,S\left(  x,y\right)  \right]  $= $x\sqcup y$. Conversely, $y\in
x\sqcup y$= $\left[  x\vee y,S\left(  x,y\right)  \right]  \Rightarrow$ $x\vee
y\leq y\Rightarrow$ $x\vee y=y\Rightarrow$ $x\leq y.$ The second part of 3 is
proved dually.
\end{proof}

\begin{proposition}
\label{cnt0308}For all $x,y\in X$ we have: $\left(  x\sqcup y\right)
^{\prime}=x^{\prime}\sqcap y^{\prime}$ and $\left(  x\sqcap y\right)
^{\prime}=x^{\prime}\sqcup y^{\prime}$.
\end{proposition}

\begin{proof}
We prove only the first part (the second part is proved dually). We have%
\begin{align*}
\left(  x\sqcup y\right)  ^{\prime}  &  =\left[  x\vee y,S\left(  x,y\right)
\right]  ^{\prime}\\
&  =\left\{  z^{\prime}:x\vee y\leq z\leq S\left(  x,y\right)  \right\} \\
&  =\left\{  z^{\prime}:x^{\prime}\wedge y^{\prime}\geq z^{\prime}\geq\left(
S\left(  x,y\right)  \right)  ^{\prime}\right\} \\
&  =x^{\prime}\sqcap y^{\prime}.
\end{align*}
\end{proof}

\begin{proposition}
\label{cnt0309}For all $x,y,z\in X$ we have:

\begin{enumerate}
\item $\left[  T\left(  x,y\vee z\right)  ,x\wedge\left(  y\vee z\right)
\right]  \subseteq\left(  x\sqcap\left(  y\sqcup z\right)  \right)
\cap\left(  \left(  x\sqcap y\right)  \sqcup\left(  x\sqcap z\right)  \right)
$.

\item $\left[  x\vee\left(  y\wedge z\right)  ,S\left(  x,y\wedge z\right)
\right]  \subseteq\left(  x\sqcup\left(  y\sqcap z\right)  \right)
\cap\left(  \left(  x\sqcup y\right)  \sqcap\left(  x\sqcup z\right)  \right)
$.
\end{enumerate}
\end{proposition}

\begin{proof}
We prove only 1 (2 is proved dually). We have%
\begin{equation}
x\sqcap\left(  y\sqcup z\right)  =x\sqcap\left[  y\vee z,S\left(  y,z\right)
\right]  =\left[  T\left(  x,y\vee z\right)  ,x\wedge S\left(  y,z\right)
\right]  . \label{eq0321}%
\end{equation}
But $x\wedge\left(  y\vee z\right)  \leq x\wedge S\left(  y,z\right)  \ $ and
so%
\begin{equation}
\left[  T\left(  x,y\vee z\right)  ,x\wedge\left(  y\vee z\right)  \right]
\subseteq\left[  T\left(  x,y\vee z\right)  ,x\wedge S\left(  y,z\right)
\right]  \label{eq0322}%
\end{equation}
Also,
\begin{equation}
\left(  x\sqcap y\right)  \sqcup\left(  x\sqcap z\right)  =\left[  T\left(
x,y\right)  ,x\wedge y\right]  \sqcup\left[  T\left(  x,z\right)  ,x\wedge
z\right]  =\left[  T\left(  x,y\right)  \vee T\left(  x,z\right)  ,S\left(
x\wedge y,x\wedge z\right)  \right]  . \label{eq0323}%
\end{equation}
But $T\left(  x,y\vee z\right)  $= $T\left(  x,y\right)  \vee T\left(
x,z\right)  $. Also $x\wedge\left(  y\vee z\right)  $= $\left(  x\wedge
y\right)  \vee\left(  x\wedge z\right)  \leq$ $S\left(  x\wedge y,x\wedge
z\right)  $. Hence
\begin{equation}
\left[  T\left(  x,y\vee z\right)  ,x\wedge\left(  y\vee z\right)  \right]
\subseteq\left[  T\left(  x,y\right)  \vee T\left(  x,z\right)  ,S\left(
x\wedge y,x\wedge z\right)  \right]  . \label{eq0324}%
\end{equation}
Now 1 follows immediately from (\ref{eq0322}) amd (\ref{eq0324}).
\end{proof}

\begin{proposition}
\label{cnt0310}The hyperalgebra $\left(  X,\sqcap,\sqcup\right)  $ is a superlattice.
\end{proposition}

\begin{proof}
The proof consists in checking that all the properties listed in Definition
\ref{cnt0213}\ are satisfied when we use $\sqcap$ in place of $\bigtriangleup$
and $\sqcup$ in place of $\bigtriangledown$. Indeed, \textbf{A1}, \textbf{A4}
and \textbf{A5} are parts 1, 2 and 3 of Proposition \ref{cnt0307} and
\textbf{A2}, \textbf{A3} are parts 2 and 4 of Proposition \ref{cnt0306}.
\end{proof}

\section{Generalizations}

\label{sec04}

We can generalize the construction of the multi-valued connectives (Definition
\ref{cnt0304}) in the following manner. Suppose that $T_{1},T_{2}$ are t-norms
and $S_{1},S_{2}$ their dual t-conorms. Furthermore, suppose that for all
$x,y\in X$ we have $T_{1}\left(  x,y\right)  \leq T_{2}\left(  x,y\right)  $
and $S_{2}\left(  x,y\right)  \leq S_{1}\left(  x,y\right)  $. For all $x,y\in
X$ define
\begin{equation}
x\sqcap y=\left[  T_{1}\left(  x,y\right)  ,T_{2}\left(  x,y\right)  \right]
,x\sqcup y=\left[  S_{2}\left(  x,y\right)  ,S_{1}\left(  x,y\right)  \right]
. \label{eq0403a}%
\end{equation}
Then it is still possible that $\sqcap$, $\sqcup$ have the t-norm, t-conorm
properties of Proposition \ref{cnt0306}. As an example take
\[
T_{1}\left(  x,y\right)  =\max\left(  0,x+y-1\right)  ,\quad T_{2}\left(
x,y\right)  =xy,\quad S_{1}\left(  x,y\right)  =\min\left(  1,x+y\right)
,\qquad S_{2}\left(  x,y\right)  =x+y-xy.
\]
It is easy to check that all the properties of Proposition \ref{cnt0306} still hold.

However, an additional attractive point of our construction is that $\left(
X,\sqcap,\sqcup\right)  $ behaves similarly to a lattice (i.e. it is a
superlattice). Can we obtain this behavior for $T_{2}$ different from $\wedge$
and $S_{2}$ different from $\vee$? A first answer turns out to be negative.

\begin{proposition}
\label{cnt0401}Suppose that $T_{1},T_{2}$ are t-norms and $S_{1},S_{2}$ their
dual t-conorms. Furthermore, suppose that for all $x,y\in X$ we have
$T_{1}\left(  x,y\right)  \leq T_{2}\left(  x,y\right)  $ and $S_{2}\left(
x,y\right)  \leq S_{1}\left(  x,y\right)  $. For all $x,y\in X$ define
$x\sqcap y$ and $x\sqcup y$ as in (\ref{eq0403a})$.$Then%
\begin{align}
\left(  \forall x,y\in X:T_{2}\left(  x,y\right)  =x\wedge y\right)   &
\Leftrightarrow\left(  \forall x,y\in X:x\leq y\Leftrightarrow x\in x\sqcap
y\right) \label{eq0401}\\
\left(  \forall x,y\in X:S_{2}\left(  x,y\right)  =x\vee y\right)   &
\Leftrightarrow\left(  \forall x,y\in X:x\leq y\Leftrightarrow y\in x\sqcup
y\right)  \label{eq0402}%
\end{align}
\end{proposition}

\begin{proof}
We have already proved that $\left(  \forall x,y\in X:T_{2}\left(  x,y\right)
=x\wedge y\right)  $ implies $(\forall x,y\in X:$ $x\leq y$ $\Leftrightarrow$
$x\in x\sqcap y$). To go the other way, suppose that $\left(  \forall x,y\in
X:x\leq y\Leftrightarrow x\in x\sqcap y\right)  $ holds. Choose any $x,y\in
X$. Since $x\wedge y\leq y$, we must have
\[
x\wedge y\in x\sqcap y=\left[  T_{1}\left(  x\wedge y,y\right)  ,T_{2}\left(
x\wedge y,y\right)  \right]  \subseteq\left[  T_{1}\left(  x\wedge y,y\right)
,T_{2}\left(  x,y\right)  \right]  .
\]
Hence $x\wedge y\leq T_{2}\left(  x,y\right)  $; but also $T_{2}\left(
x,y\right)  \leq x\wedge y$. It follows that $T_{2}\left(  x,y\right)
=x\wedge y$ and this holds for every $x,y\in X$. Hence (\ref{eq0401}) has been
proved. (\ref{eq0402})\ is proved dually.
\end{proof}

From the above proposition we see that $\left(  X,\sqcap,\sqcup\right)  $ is a
superlattice \emph{compatible with the original order }$\leq$ iff $x\sqcap y$
and $x\sqcup y$ are defined according to Definition \ref{cnt0304}.

However it may still be possible to define $x\sqcap y$ and $x\sqcup y$ in such
a manner that $\left(  X,\sqcap,\sqcup\right)  $ is a superlattice in a more
general sense. Namely, suppose that \textbf{A1-A4} are satisfied and
\textbf{A5} is replaced by the following conditions.

\begin{description}
\item[A6] $y\in x\bigtriangledown y$ $\Leftrightarrow x\in x\bigtriangleup y$.

\item[A7] $\left(  x\in x\bigtriangledown y\text{ and }y\in x\bigtriangledown
y\right)  \Rightarrow x=y$.

\item[A8] $\left(  x\in x\bigtriangledown y\text{ and }y\in y\bigtriangledown
z\right)  \Rightarrow x\in x\bigtriangledown z$.
\end{description}

If \textbf{A1-A4} and \textbf{A6-A8} hold, then we can define a relation
$\leqslant$ on $X$ as follows: ``$x\leqslant y$ iff $y\in x\bigtriangledown
y$''. It turns out that using \textbf{A6-A8} it can be shown that $\leqslant$
is an order on $X$, which will, in general, be different from $\leq$; in fact
\textbf{A1-A4} and \textbf{A6-A8} do not use $\leq$ at all, hence the
hyperoperations $\bigtriangledown,\bigtriangleup$ can be defined in a general
set $X$ (not necessarily a lattice).

In this light, it may be possible for some pairs $T_{1}$, $T_{2}$ and $S_{1}$,
$S_{2}$ to define $\sqcup$ and $\sqcap$ as in (\ref{eq0403a}) and then show
that \textbf{A1-A4} and \textbf{A6-A8} hold; in such a case $\sqcup$ and
$\sqcap$ will define an order $x\leqslant y$ on $X$ as follows: ``$x\leqslant
y$ iff $y\in x\sqcup y$'' and $\sqcup$, $\sqcap$ are reasonable candidates for
multi-valued t-norm and t-conorm on $X$. However, we emphasize again that
$\sqcup$, $\sqcap$ will not fully respect the ``intrinsic'' order $\leq$.

\section{Conclusion}

\label{sec05}

We have presented a procedure for constructing \emph{multi-valued} t-norms and
t-conorms. Our construction uses a pair of single-valued t-norms and the pair
of dual t-conorms and constructs interval-valued t-norms $\sqcap$ and
t-conorms $\sqcup$. In this manner we can combine desirable characteristics of
different t-norms and t-conorms; furthermore if we use the t-norm $\wedge$ and
t-conorm $\vee$, then $\left(  X,\sqcap,\sqcup\right)  $ is a superlattice,
i.e. the multivalued analog of a lattice.

Let us close with some issues which require further research. First, it will
be interesting to obtain further ``deMorgan-like'' properties of $\left(
X,\sqcap,\sqcup,^{\prime}\right)  $ and develop a logic based on multi-valued
connectives. Of particular interest is the study of the resulting implication
operator, the law of excluded middle and the law of contradiction. Second,
note that the fuzzy implication operator is closely connected to the
\emph{fuzzy inclusion measure}, so it would be interesting to consider
interval-valued inclusion measures. Third, we are interested in analyzing
$\left(  X,\sqcap,\sqcup\right)  $ from a geometric point of view, paying
special attention to issues such as metric properties, continuity, convexity
and betweenness. Finally, it will be interesting to develop a procedure for
developing a \emph{family }of interval-valued t-norms $\left\{  \sqcap
_{a}\right\}  _{a\in\left[  0,1\right]  }$ which have the $a$\emph{-cut
properties}, because the $\sqcap_{a}$'s can then be used to construct a
\emph{fuzzy-valued t-norm }$\overline{\sqcap}$. Similarly, one could use a
family $\left\{  \sqcup_{a}\right\}  _{a\in\left[  0,1\right]  }$ to construct
a \emph{fuzzy-valued t-conorm }\underline{$\sqcup$}.

\end{document}